\documentclass[12pt,a4paper]{article}

\usepackage[latin1]{inputenc}
\usepackage{amsmath}
\usepackage{amsfonts}
\usepackage{amssymb}
\usepackage{epsfig}
\usepackage{graphics}
\usepackage{graphicx}
\newcommand{\ba}{\begin{array}}
\newcommand{\ea}{\end{array}}
\newcommand{\bear}{\begin{eqnarray}}
\newcommand{\ear}{\end{eqnarray}}

\def\be{\begin{equation}}
\def\ee{\end{equation}}
\def\ds{\displaystyle}
\def\bea{\begin{equation} \begin{array}}
\def\eea{\end{array} \end{equation}}
\def\bal{\begin{equation} \begin{aligned}}
\def\eal{\end{aligned} \end{equation}}

\begin{document}

\begin{titlepage}

\vspace*{1.8cm}

\begin{center}
{\bf\Large Light fermion masses and chiral freedom}

\vspace{1cm}
C. Wetterich\\
\bigskip
Institut  f\"ur Theoretische Physik\\
Universit\"at Heidelberg\\
Philosophenweg 16, D-69120 Heidelberg\\

\vspace{0.5cm}
\end{center}

\begin{abstract}
Within the scenario of chiral freedom we compute the quark and lepton masses of the first two generations in terms of their chiral couplings. This allows us to make a rough estimate of the size of the chiral couplings, narrowing down the uncertainty in the chiron contribution to low energy observables, like the anomalous magnetic moment of the muon. We also extract information about the chiron mass which determines the size of possible chiron effects at the LHC.

\end{abstract}
\end{titlepage}

Chiral freedom has been proposed \cite{CW1} \cite{CW3} as a possible solution to the gauge hierarchy problem. In a theory with chiral antisymmetric tensor fields a local mass term can be forbidden by a symmetry. The Fermi-scale of the weak interactions is then generated by dimensional transmutation. The chiral couplings between the quarks and the antisymmetric tensor fields - the chirons - are asymptotically free. Similar to the running gauge coupling in QCD they generate a non-perturbative mass scale where they grow large. If such models are viable, the LHC-experiments may be able to test this alternative to the standard model with a ``fundamental'' Higgs scalar.

The consistency of models with chiral antisymmetric tensors depends on properties of the ground state \cite{CW2}. In particular, the generation of a non-local mass term seems to be needed. This issue is difficult to settle since it concerns the properties of the chiron propagator in a momentum range where the chiral couplings are large and non-perturbative. In the meanwhile, one may ask if such models are compatible with the present precision tests of electroweak interactions. Computations of the anomalous magnetic moment of the muon, the LEP-precision tests or interactions contributing to the b-\={b}-forward-backward asymmetry all show \cite{CW3} a strong dependence on the unknown size of the chiral couplings to muons or b-quarks, as well as to the infrared properties of the chiron propagator, e.g. the non-local chiron mass.

In this note we attempt a rough estimate of these unknown parameters. For this purpose we use the fact that the chiral coupling to the muon, $f_\mu$, determines the size of the muon mass $m_\mu$ - and similar for the other ``light'' generations. We will see how a computation of the ratio $m_\mu / f_\mu$ will give  an estimate of $f_\mu f_b$. This allows for a quantitative relation between the muon anomalous magnetic moment $g-2$ and the low momentum behavior of the chiron propagator. Even though the accuracy of these computations is limited by non-perturbative physics, the overall magnitude of some of the couplings can be constrained so that first semi-quantitative estimates of the size of possible deviations form the standard model become possible.

The chiral couplings of the antisymmetric tensor fields $\beta^\pm_{\mu\nu}$ to the quarks and leptons are given by \cite{CW3}, \cite{CH}
\begin{eqnarray}\label{A}
-{\cal L}_{ch}&=&\bar{u}_R\bar{F}_U\tilde{\beta}_+q_L-\bar{q}_L
\bar{F}^\dagger_U\stackrel{\eqsim}{\beta}_+u_R\nonumber\\
&&+\bar{d}_R\bar{F}_D\bar{\beta}_-q_L-\bar{q}_L\bar{F}^\dagger_D\beta_-d_R\nonumber\\
&&+\bar{e}_R\bar{F}_L\bar{\beta}_-l_L-\bar{l}_L\bar{F}^\dagger_L\beta_-e_R,
\end{eqnarray}
with
\be\label{B}
\beta_\pm=\frac12\beta^\pm_{\mu\nu}\sigma^{\mu\nu}~,~\sigma^{\mu\nu}=\frac i2 [\gamma^\mu,\gamma^\nu].
\ee
With respect to the electroweak gauge symmetry the chirons $\beta^\pm$ transform as doublets with hypercharge $Y=1$, similar to the Higgs doublets. (For details see \cite{CW3}.) The model has a discrete symmetry $\beta^-\to -\beta^-$. 

The chiral couplings are organized in $3\times 3$-matrices $F_{U,D,L}$. Similar to the Yukawa couplings in the standard model they carry the information about (approximate) flavor symmetries and mixings. Indeed, we can choose phase conventions for the fermions such that $F_{U,D,L}$ are diagonal and real. All information on flavor mixing and CP-violating phases appears then only in the interactions of the $W$-boson in the form of the usual CKM-matrix \cite{CKM}. These effects are small and will be neglected for the purpose of this note. In this limit separate conserved flavor symmetries are present, like a conserved $L_\mu$ for the lepton number of muons and muon-neutrinos. This implies that also the mass matrices for the quarks and leptons must be real and diagonal. We will choose conventions where all mass eigenvalues are positive. With these conventions the sign of the chiral couplings matters - indeed the sign of $m_\mu/ f_\mu$ is, in principle, measurable and cannot be changed by chiral phase rotations. For example, this sign will determine the sign of $g-2$.

In the limit of vanishing $f_\mu$ the model exhibits an additional axial symmetry for the muons. (Effects of axial anomalies are negligible for this note.) In turn, this implies a vanishing $m_\mu$. We conclude $m_\mu= hf_\mu$,  where the proportionality constant $h$ depends on the various couplings of the model and needs to be computed. Since $h$ can only involve even powers of $f_\mu$ we will restrict the computation of $m_\mu$ to the order linear in $f_\mu$. Of course, this is only justified if $f_\mu$ turns out to be small enough. We will present our computation in the form of a loop calculation. We will argue below that this form is indeed justified even in a non-perturbative context, if appropriate effective couplings and propagators are used. The loop calculation should therefore give a reasonable estimate of the orders of magnitude involved. 

\begin{figure}[htb]
\centering
\includegraphics[scale=1.0]{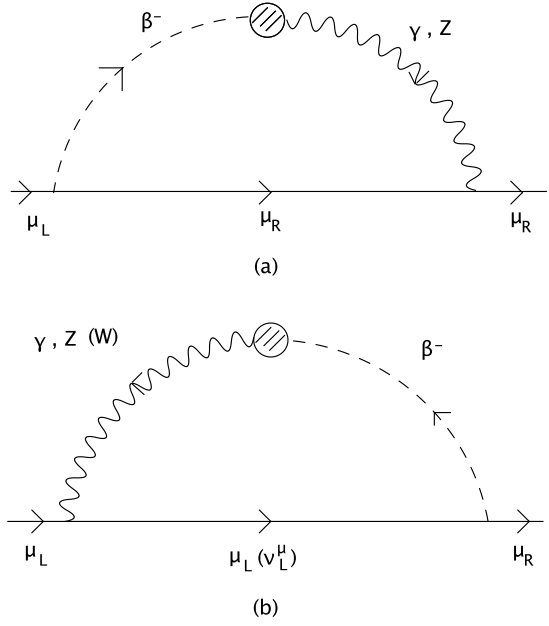}
\caption{Loop contributions to the muon mass}
\label{light-fig1}
\end{figure}
\noindent

\begin{figure}[htb]
\centering
\includegraphics[scale=1.0]{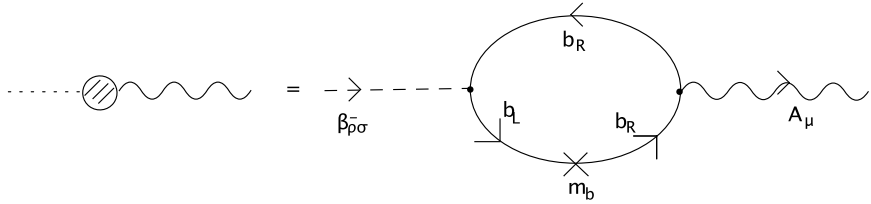}
\caption{Chiron-photon mixing}
\label{light-fig2}
\end{figure}

In linear order in the chiral coupling $f_\mu$ the muon mass gets contributions from the diagrams shown in fig. \ref{light-fig1}. In turn, the effective mixing between $\beta^-$ and the gauge bosons $\gamma,Z,W^\pm$ involves a loop with the exchange of a virtual $b$-quark, as shown in fig. \ref{light-fig2}. Thus the muon mass arises effectively in two loop order. Parametrically one finds
\be\label{eq.1}
m_\mu=c_\mu f_\mu f_b e^2 m_b
\ee
and our task is the computation of the proportionality coefficient $c_\mu$. One expects that $c_\mu$ depends on the chiron mass $m_-$ and the couplings, masses and mixings for the $W$ and $Z$ bosons.

We start with the contribution involving a photon line in the loop. It is convenient to organize the computation in two steps. We first compute the effective momentum dependent chiral vertex of the photon to the charged leptons and down quarks
\bear\label{eq.2}
\Gamma_{\gamma,\text{ch1}} 
\ds & = &\int_p\int_q\alpha_\gamma(p,q)\left\{\bar{e}_R(q+p)F_L\sigma^{\mu\nu}e_L(q)\right.\nonumber\\
&&\left.+\bar{d}_R(q+p)F_D\sigma^{\mu\nu} d_L(q)\right\}F_{\mu\nu}(p) + \text{h.\,c.}
\ear
Here $e_R=(e_R,\mu_R,\tau_R)$ is a vector with three flavor components for the right handed leptons and similar for $e_L,d_R$ and $d_L$. The $3\times 3$ matrices $F_L,F_D$ denote the renormalized chiral couplings to the antisymmetric tensors and we may include their momentum dependence if needed. For our specific case of the muon we replace $\bar{e}_RF_L\sigma^{\mu\nu}e_L$ by $f_\mu\bar{\mu}_R\sigma^{\mu\nu}\mu_L$.

\begin{figure}[htb]
\centering
\includegraphics[scale=1.0]{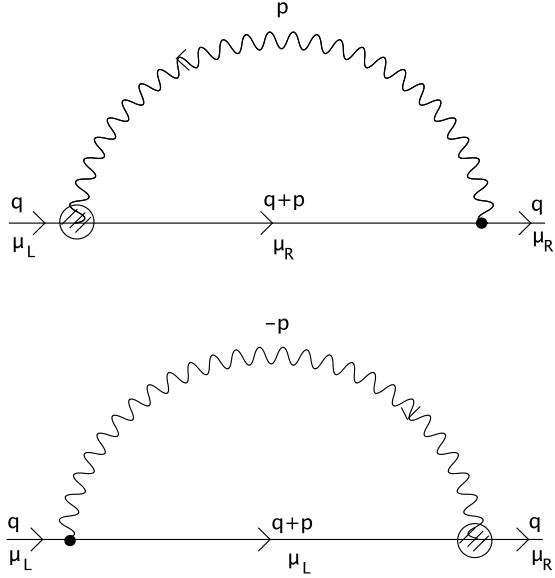}
\caption{Loops with effective chiral photon vertex}
\label{light-fig3}
\end{figure}

Secondly, we evaluate the diagrams in fig. \ref{light-fig3}, where the circles denote the effective chiral photon vertex (\ref{B}).

\noindent
This yields, for $q^2\to0$ and with $\sigma_{\mu\nu} \sigma^{\nu\mu}= -6$, $f_\mu=f_\mu(p=0)$,
\bea{r@{\;=\;}l} \label{eq.3}
\ds m_\mu^{(\gamma)} & \ds 2e \int_p f_{\mu}(p) \alpha_\gamma (p,0) \frac{p_\mu p_\rho}{p^4} \left\{\sigma^{\mu\nu} \gamma^\rho \gamma_\nu - \gamma_\nu \gamma^\rho \sigma^{\mu\nu}
\right\}\\[4mm]
&\ds -6ief_\mu \int_p \frac{\alpha_\gamma (p,0)}{p^2} \frac{f_\mu(p)}{f_\mu} \;.\\[6mm]
\eea

\begin{figure}[h!tb]
\centering
\includegraphics[scale=1.2]{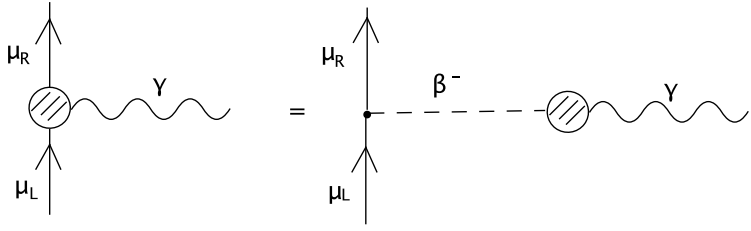}
\caption{Effective chiral vertex of the photon}
\label{light-fig4}
\end{figure}

In lowest order in the mixing between the chiral tensors and the gauge bosons the effective chiral vertex of the photon is composed of the vertex for the coupling of the chiral tensor to the muon $(\sim f_\mu \sigma^{\mu\nu})$, the mixing between the chiral tensor and the photon $\epsilon^-_\gamma$, and the propagator of the chiral tensor, as shown in fig.
\ref{light-fig4}. The dominant momentum dependence of $\alpha_\gamma (p,q=0)=\alpha_\gamma (p)$ arises from the propagator of the chiral tensor field $B^{-0}$
\be \label{eq.4}
\alpha_\gamma (p) = - \frac{\epsilon^-_\gamma (p)}{2 \sqrt{2}} \frac{c_{\beta,\gamma}}{Z_-(p)p^2+m^2_-(p)} \;.
\ee

\noindent
For simplicity we neglect the mixing of the real part of the neutral component of $\beta$, i.e. $(\beta^{-0}+(\beta^{-0})^*) /\sqrt{2}$, with other neutral components of the chiral tensor or with the Z-boson. In this approximation one has $c_{\beta,\gamma} =1$, whereas more generally the additional mixing effects can be incorporated in $c_{\beta,\gamma}$. 

The photon-chiron mixing coefficient $\epsilon_\gamma^- (p)$ corresponds to a piece in the effective action 
\be\label{eq.5}
\Gamma_m=-\frac{1}{2 \sqrt{2}} \int_p \epsilon^-_\gamma (p) \left[ \beta^{-0}_{\mu\nu}(-p)+\beta^{-0}_{\mu\nu}(p)^* \right]F^{\mu\nu}(p).
\ee
One finds
\be\label{eq.6}
\epsilon^-_\gamma(p)=\frac{8ie}{\sqrt 2} m_b(p) f_b(p) I_1 (p)
\ee
where we approximate the momentum integral in the loop similar to
\cite{CW3}

\bear\label{eq.7}
I_1(p)&\approx&\int\limits_q\left[\left(q+\frac{p}{2}\right)^2+m^2_b\right]^{-1}
\left[\left(q-\frac p2\right)^2+m^2_b\right]^{-1}\nonumber\\
&\approx& \frac{i}{16\pi^2} \ln \left( \frac{\Lambda^2_b+p^2}{m^2_b+p^2/8} \right).
\ear
(For this approximation to $I_1$ we have restricted the euclidean momentum integral in the b-quark loop (fig. \ref{light-fig2})
to $q^2 < \Lambda^2_b+p^2$, taking into account that $m_b(q+p)/m_b(p)$ vanishes for large loop momenta $q$.) Combining the different pieces yields
\be\label{eq.8}
\alpha_\gamma(p)= \frac{c_{\beta,\gamma}}{8\pi^2} \frac{e\, m_b(p)f_b(p)}{Z_-(p) p^2 +m_-^2 (p)} \ln \frac{\Lambda_b^2 +p^2}{m^2_b+p^2/8}.
\ee

We next evaluate the loop integral for $m^{(\gamma)}_\mu$ (\ref{eq.3}) by analytical continuation of $q_0$ to euclidean  signature. The momentum dependence of the logarithmic factor $I_1(p)$ in $\alpha_\gamma(p)$ is weak and we will replace $I_1(p)$ by a constant $I_1(p^2=m^2_-)$. For a computation of $m_\mu^{(\gamma)}$, cf. eq. \eqref{eq.3}, we therefore need to evaluate the integral
\be\label{eq.8a}
I_2=i \int\limits_p^{(E)} \frac{1}{p^2(Z_-(p)p^2+m^2_-(p))} \frac{f_\mu(p)}{f_\mu} \frac{f_b(p)}{f_b}  \frac{m_b(p)}{m_b}
\ee
with $f_{\mu,b}=f_{\mu,b}(p=0)$, $m_b=m_b(0)$. Here $\int^{(E)}_p =\int^{(E)} \frac{d^4p}{(2\pi)^4}$ denotes the integral with euclidean signature. The precise meaning of the wave function renormalization $Z_-{(p)}$ and the couplings $f_{\mu,b}(p)$ depends on the normalization of the chiral tensor field. It is convenient to adopt a different normalization for large and small $p^2$. For the range $p^2 < m^2_-$ we work in a momentum independent normalization for the chiral tensor field, where $Z_-(p^2<m^2_-)$ may depend on $p$,  with $m^2_-=m^2_-(p^2=m^2_-)$. For $p^2 > m^2_-$ we adopt a momentum dependent normalization, keeping the  `kinetic term' $\sim p^2$ fixed in the inverse propagator, $Z_-(p^2\ge m^2_-)=1$. 

The momentum dependence of $f_b(p)$ and $m_b(p)$ will ensure the ultraviolet finiteness of $I_2$. Let us first consider the range $m_b^2<p^2<m_-^2$. For a momentum independent normalization of the chiral tensors the effects from the anomalous dimension of these fields should not be included in the momentum-dependence of the chiral couplings. Then the chiral couplings $f_{\mu,b}(p)$ are essentially constant for $p^2<m^2_-$ and we may associate $f_{\mu,b}=f_{\mu,b}(p^2=m^2_-)$. With this convention the anomalous dimension induced by b-quark loops shows only up in the kinetic term for the chiral tensors
\be\label{eq.8b}
Z_- (p)= \frac{f_b^2+\frac13 f^2_\tau}{4\pi^2} \ln \frac{m^2_-}{p^2+m^2_b} \quad \text{for} \quad p^2<m^2_- \;.
\ee
Since for $p^2\ll m_-^2$ the inverse chiron-propagator is dominated by $m_-^2$ the momentum dependence of $Z_-(p)$ is actually not very important. We neglect it and set $Z_-=1$.

On the other hand, for $p^2>m_-^2$ the inverse chiron propagator is dominated by $p^2$ and the momentum dependence of $m^2_-(p)$ becomes  unimportant. We may effectively replace $m^2_-(p)$ by the constant $m^2_-$ which corresponds to a suitable average value of $m^2_-(p)$ for $p^2\approx m^2_-$. For $p^2>m_-^2$ the chiral coupling of the muon is enhanced
\be\label{eq.8c}
\begin{aligned}
\frac{f_\mu(p)}{f_\mu} &= \left(\frac{p^2}{m_-^2}\right)^{\frac{\bar f^2 (p)}{8\pi^2}}\;,\\
\bar f^2(p) &= \int\limits^{\ln|p|}_{\ln m_-} \!\!\! dt\, \left(f^2_b(t)+\frac13 f^2_\tau (t)\right) \Big/ \ln \frac{|p|}{m_-} \;.
\end{aligned}
\ee
We observe that $f_b(p)$ decreases much faster such that the factor $f_\mu (p) f_b(p)$ decreases for increasing $p^2$. Furthermore, $m_b(p)$ is expected to decrease and to vanish for sufficiently large $p^2$. We combine these effects in an effective ultraviolet cutoff, restricting the integral $I_2$ to $p^2<\tilde\Lambda ^2$, and approximate
\be \label{eq.9}
I_2=\frac{i}{16 \pi^2} \ln \frac{\tilde\Lambda^2+m^2_-}{m^2_-} \;.
\ee

Combining the different pieces yields
\be \label{eq.9A}
\ds m_\mu^{(\gamma)}=\frac{3c_{\beta,\gamma}}{64\pi^4}  \ln \left[8 \left(1+\frac{\Lambda^2_b}{m^2_-}\right) \right] \ln \left( 1 +\frac{\tilde\Lambda^2}{m_-^2} \right) e^2 f_b f_\mu m_b \;.
\ee
Our rather crude discussion reveals that this result still has substantial uncertainties, cast here into the form of effective cutoffs $\Lambda_b$ and $\tilde\Lambda$ and the mixing coefficient $c_{\beta,\gamma}$. Nevertheless, $c_{\beta,\gamma}$ is expected to be of the order of magnitude one. The dependence on $\Lambda_b$ and $\tilde\Lambda$ is only logarithmic.

It is useful to trace the origin of the different factors appearing in the ratio $m_\mu/m_b$. For this  purpose we introduce the coefficient
\be \label{eq.9B}
\kappa_\mu = \frac{m_\mu}{m_\mu^{(\gamma)}}
\ee
for the ratio between the total muon mass and the contribution from the photon loop. We write 
\be
\begin{aligned} \label{eq.10}
m_\mu &= \kappa_\mu e\,  f_\mu \hat\alpha_\gamma b_\mu \;,\\
\hat\alpha_\gamma &= \frac{c_{\beta,\gamma}}{8\pi^2} e f_b m_b \ln \left[ 8 \left(1+ \frac{\Lambda^2_b}{m^2_-}\right)  \right] \;,\\
b_\mu &= \frac{3}{8\pi^2}  \ln \left( 1+\frac{\tilde\Lambda^2}{m_-^2} \right),
\end{aligned}
\ee
which separates the factors arising from the chiral coupling of the photon ($\hat\alpha_\gamma$) and the remaining loop integral ($b_\mu$). Here $f_\mu\hat\alpha_\gamma/(m^2_-(p)+Z_-p^2)$ measures the strength of the anomalous muon-photon coupling $f_\mu(p)\alpha_\gamma(p)$, suitably averaged in the relevant momentum range. Since the coefficient $b_\mu$ is positive, the sign of $f_\mu\hat\alpha_\gamma/m_\mu=(eb_\mu\kappa_\mu)^{-1}$ depends only on the sign of $\kappa_\mu$. This sign will determine the sign of the correction $\Delta(g-2)$ to the anomalous magnetic moment of the muon. 
In terms of the factors \eqref{eq.10} the coefficient $c_\mu$, in eq. \eqref{eq.1}, reads
\be
c_\mu = \frac{\kappa_\mu b_\mu \hat\alpha_\gamma}{e f_b m_b} \;.
\ee
\\

What remains is the computation of $\kappa_\mu$. For this purpose we have to compare the weight of the diagrams involving the exchange of $Z$-or $W$-bosons with the photon exchange diagram. The effective coupling of the $Z$-boson to the muon has the same chiral structure as for the photon. We only have to replace the photon-chiron mixing by the $Z$-chiron mixing. Working again in the approximation where $(\beta^{-0}+\beta^{-0*})/\sqrt{2}$ dominates, we find a relative factor
\be\label{a5a}
f_{Z,b}=\frac{1}{\cos\vartheta_W\sin\vartheta_W}\left(\frac34-\sin^2\vartheta_W\right)
\ee
from the coupling of the $Z$-boson to the $b$-quark in the diagram corresponding to fig. 2. (Only the vector coupling of $Z$ contributes, not the axial vector.) A similar factor arises from the coupling to the muon (again only the vector coupling matters)
\be\label{a5b}
f_{Z,\mu}=\frac{3}{4\cos\vartheta_W\sin\vartheta_W}(1-4\sin^2\vartheta_W).
\ee

For the diagrams involving the exchange of the $W$-boson we first evaluate the effective chiral vertex of the $W$-boson with the leptons $(W^\mu=W^{+\mu}=(W^\mu_1-iW^\mu_2)/\sqrt{2},~W_{\mu\nu}=\partial_\mu W_\nu-\partial_\nu W_\mu)$
\be\label{a5c}
\Gamma_{W,ch1}=\int\limits_p\int\limits_q\alpha_W(p,q)\bar e_R(q+p)F_L\sigma^{\mu\nu}\nu_L(q)W_{\mu\nu}(p)+h.c.
\ee
A computation similar to the chiron-photon mixing yields
\bear\label{a5d}
\alpha_W(p,q)&=&\alpha_\gamma(p,q)f_{W,b}\frac{c_{\beta,W}}{c_{\beta,\gamma}}~,\nonumber\\
f_{W,b}&=&-\frac{3}{2\sqrt{2}\sin\vartheta_W}.
\ear
Here $f_{W,b}$ accounts for the different coupling of the $W$-boson to the $b$-quark as compared to the photon. The factor $c_{\beta,W}$ accounts again for mixing effects in the chiron propagator and incorporates the possibility that the mass of the charged chiron $\beta^{-+}$ may differ from the neutral chiron $\beta^{-0}$.  In the $W$-exchange diagram in fig. 1b we pick up an additional factor $f_{W,\mu}$ from the $W{\mu\bar\nu}$ vertex 
\be\label{a5e}
f_{W,\mu}=-\frac{1}{2\sqrt{2}\sin\vartheta_W}.
\ee
Summing up all diagrams this yields for $\kappa_\mu$
\be\label{a5f}
\kappa_\mu=1+f_{Z,b}f_{Z,\mu}\frac{c_{\beta,Z}}{c_{\beta,\gamma}}+f_{W,b}f_{W,\mu}
\frac{c_{\beta,W}}{c_{\beta,\gamma}}.
\ee
If we assume for a moment $c_{\beta,Z}=c_{\beta,W}=c_{\beta,\gamma}$ we obtain
\be\label{a5g}
\bar\kappa_\mu=\frac{11-33\sin^2\vartheta_W+16\sin^4\vartheta_W}{8\cos^2\vartheta_W\sin^2\vartheta_W}\approx 3.5
\ee
and find a moderate enhancement as compared to the pure photon-exchange contribution. This is mainly due to the dominant $W$-exchange diagram. Since the $W$-exchange contribution can easily overweigh the photon exchange, an assessment of the sign of $\kappa_\mu$ requires an investigation of the possible sign of $c_{\beta,W}/c_{\beta,\gamma}$. We will turn to this question below. 

An important ingredient for our computation of $m_\mu/m_b$, namely the effective chiral coupling of the photon (\ref{eq.2}), is accessible to measurement for low momenta $p^2$. Indeed, we note an interesting relation between our computation of the relation between $m_\mu$ and $f_\mu$ on the one side and a contribution to the anomalous magnetic moment of the muon on the other side. The chiral coupling of the photon in eq. \eqref{eq.2} directly contributes to $g-2$ \cite{CW3}
\be \label{eq.A}
\Delta(g-2) = - \frac{8m_\mu}{e} f_\mu \alpha_\gamma (0) \; .
\ee
Using the experimental bounds on $\Delta(g-2)$ this will allow us to restrict the allowed range for certain parameters. In particular, the product $f_\mu f_b$ can be eliminated in favor of $\Delta(g-2)$. In order to exploit the relation (\ref{eq.A}) we may write eq. \eqref{eq.10} as
\be \label{eq.B}
m_\mu = ef_\mu \alpha_\gamma (0) m_-^2 (0)H\kappa_\mu
\ee
with
\be \label{eq.C}
H= \frac{\hat\alpha _\gamma}{\alpha_\gamma(0)m_-^2(0)} b_\mu =K b_\mu.
\ee

Here we recall that $\hat\alpha_\gamma$ is essentially the chiral photon coupling with the  chiron propagator amputated and  evaluated for an appropriate momentum $p^2\approx m^2_-$. The factor $K$ therefore accounts for the momentum dependence of the effective anomalous photon vertex. We write it in the form
\be\label{c5a}
K=\frac{I_1(p^2=m^2_-)}{I_1(p^2=0)}\tilde{f}_K=
\frac{\ln 8\left[1+{\Lambda^2_b}/{m^2_-}\right]}{\ln[\Lambda^2_b/m^2_-)}f_K
\ee
where the second equation uses our simple estimates of the relevant momentum integrals, cf. eq. \eqref{eq.7}. The factor $f_K$ contains the uncertainties in this estimate and the possibility of a momentum dependence of $c_{\beta,\gamma}$. We will take $f_K=1$. In other words, the factor $K$ reflects the fact that $\Delta(g-2)$ measures $\alpha_\gamma(p^2)$ for $p^2\to m^2_\mu$ where the effective chiron propagator is constant $\sim c_{\beta,\gamma}(0)/m^2_-(0)$. In contrast, the momentum integral in the loop in fig. 2 covers effectively the range $m^2_b<q^2<\Lambda^2_b$. Here the integral for $\alpha_\gamma(p^2\gg m^2_b)$ is cut off by $p^2/8$. One therefore expects $K<1$ and we will take $K\approx 0.3$.

We can now use eq. \eqref{eq.B} in order to eliminate $f_\mu \alpha_\gamma (0)$ in eq. \eqref{eq.A}
\be \label{eq.D}
\Delta(g-2)=-\frac{8m_\mu^2}{H\kappa_\mu e^2 m_-^2 (0)} \approx -2.5\cdot10^{-6}\frac{\bar H}{H}
\frac{\bar\kappa_\mu}{\kappa_\mu}
\frac{1TeV^2}{m^2_-(0)},
\ee
with $\bar H\bar\kappa_\mu=0.1$ reflecting our crude estimates. The experimental information on $\Delta(g-2)$ can directly be used for limiting the acceptable range of $m^2_-(0)$. Much of the uncertainties in the estimates of \cite{CW3}, which have involved the unknown chiral couplings $f_\mu$ and $f_b$, are now eliminated by our computation of $m_\mu/m_b$ where the same couplings enter. We may take the present difference between the experimentally measured \cite{MEX} mean value of $g-2$ and the mean value of theoretical estimates \cite{MTH} as a guide, $\Delta(g-2)=6\cdot 10^{-9}$. For $H\kappa_\mu=-\bar H\bar\kappa_\mu$ eq. \eqref{eq.D} yields $m_-{(0)}=20 TeV$. This value is considerably larger than the top quark mass. A large ratio $m_-{(0)}/m_t$ will require an explanation. Unfortunately, a large chiron mass would also render the experimental detection of effects mediated by the chiral tensors much harder. 

We may also us the relation \eqref{eq.10} for an estimate of the product of chiral couplings $f_\mu f_b$. For our crude estimate we find
\be\label{e5a}
\frac{m_\mu}{m_b}=2\cdot 10^{-3}f_\mu f_b~\frac{\kappa_\mu}{\bar\kappa_\mu},~f_\mu f_b\approx 5\frac{\bar\kappa_\mu}{\kappa_\mu}.
\ee
The rather large value of $f_\mu f_b$ is at the origin of the high value of $\bar H$ and therefore the large value of $m_-(0)$. For $f^2_b \approx 8\pi$ one finds $f^2_\mu/(4\pi)\approx 0.1$, justifying the expansion linear in $f_\mu$. 

Substantial uncertainties remain in our computation. They mainly concern two issues. One concerns the estimate of the momentum dependence of the effective chiral photon coupling. In particular, the momentum dependence of the chiron propagator for small $p^2$ can be substantial, with $m^2_-\ll m^2_-(0)$. It has been argued in \cite{CW2} that the stability of chiral tensor models requires an asymptotic vanishing of the inverse chiron propagator $m^{-2}_-(p)$ for $p^2\to 0$, at least $\sim p^2$. In this case a substantially higher value of $m^2_-(0)\widehat{=}m^2_-(p^2=m^2_\mu)$ as compared to $m^2_-\widehat{=}m^2_-(p^2=m^2_-)$ is suggested. This effect could suppress the contribution to $\Delta(g-2)$ to a very small or even practically unoberservable level, even for values $m^2_-\lesssim 1 TeV^2$. Rather strong observable chiron-effects at the LHC could be compatible with very small $\Delta(g-2)$. 

The other uncertainty arises from the relative weight of $W$- and $Z$-exchange contributions to $m_\mu/m_b$ as compared to the photon contribution, casted here in the coefficient $\kappa_\mu$. In particular, this concerns the sign of $\kappa_\mu$ and therefore the sign of $\Delta(g-2)$. Any enhancement in $H$ or $\kappa_\mu$ would result in a corresponding decrease of $m^2_-(0)$. For both cases the main uncertainty arises from the ``mixing coefficients'' $c_{\beta,\gamma}, c_{\beta, W}, c_{\beta,Z}$ which multiply the effective chiral vertices of the gauge bosons. For fixed chiral vertices the uncertainties of the evaluation of the loop diagram in fig. 3 seem only moderate, since the dependencies on effective infrared and ultraviolet cutoffs are only logarithmic.

At this point a few general remarks on the status of our loop computation in a non-perturbative setting seem appropriate. Indeed,our computation contains an essentially non-perturbative part, namely the effective chiral vertices of the gauge bosons. This is combined with the perturbative evaluation of the loop diagrams in fig. \ref{light-fig3}, resulting in eq. \eqref{eq.1}, and the factor $b_\mu$ in eq. \eqref{eq.10}. While  the uncertainties in $b_\mu$ are only logarithmic, the estimate of the coefficients $c_{\beta,\gamma}, c_{\beta,Z}$ and $c_{\beta,W}$ is very sensitive to the non-perturbative physics.

The diagrams in fig. \ref{light-fig1} are actually the only contributions to $m_\mu$ in linear order in $f_\mu$. Indeed, the conserved quantum number $L_\mu$ (in the limit of unit CKM-matrix) implies that for arbitrary diagrams the muon line connecting the in- and out-going muon must be continuous. Since for $f_\mu=0$ the muon mass vanishes, one vertex must involve the chiron and be proportional to $f_\mu\sigma^{\mu\nu}$. On the other hand, this vertex must be reconnected to the muon line. For a contribution to $\bar\mu_R\mu_L$ with chirality flip, this necessarily involves a gauge boson vertex $\sim\gamma^\mu$. Indeed, a contribution to $m_\mu$ must involve an odd number of vertices with chirality flip along the muon line. A second vertex involving a chiron would induce a further chirality flip and an additional power of $f_\mu$. Also a possible vertex with a composite scalar field would induce a further chirality flip. It would therefore have to occur in an even number. (Also an even number of further gauge boson vertices are possible. This leads to radiative corrections to the diagram of fig. \ref{light-fig1}, that may be neglected.)  

We conclude that all contributions to $m_\mu$ which are linear in $f_\mu$ must involve a mixing between chiral tensors and gauge bosons. Summing over all fields with a nonvanishing chiral vertex to the muon one arrives at the description in terms of the effective chiral vertices of the gauge bosons. The propagation and mixings of the chirons are affected strongly by non-perturbative effects, such that the effective chiral vertices of the gauge bosons and their momentum dependence are essentially non-perturbative. 

An important non-perturbative effect concerns the mixing between the antisymmetric tensors $\beta^+_{\mu\nu}$ and $\beta^-_{\mu\nu}$ which is induced by the electroweak symmetry breaking \cite{CW3}. For the charged chirons such a mixing appears in the form of an effective Lagrangian
\be\label{z4a}
-{\cal L}_{M\beta}=\frac{\tilde m^2}{4}\beta^{\mu\nu^*}_-\beta_{+\mu\nu}+h.c.
\ee

\begin{figure}[h!tb]
\centering
\includegraphics[scale=1.2]{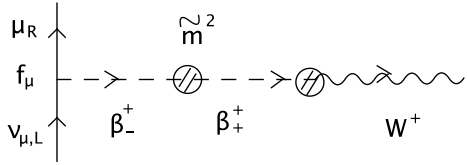}
\caption{Mixing contribution to the effective chiral vertex of the $W$-boson.}
\label{light-fig5}
\end{figure}
\noindent

\begin{figure}[h!tb]
\centering
\includegraphics[scale=1.2]{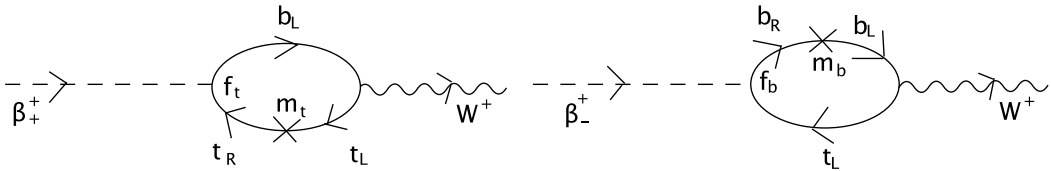}
\caption{Chiron-$W$-boson mixing}
\label{light-fig6}
\end{figure}

This generates an additional contribution to the chiral vertex of the $W$-bosons to the muon, according to fig. \ref{light-fig5}. The mixing between $\beta^+$ and $\beta^-$ is small, $\tilde m^2/(m^2_+-m^2_-)\ll 1$. The off-diagonal mass term $\tilde m^2$ involves the vacuum expectation values of the two composite Higgs scalars that are responsible for the top and bottom masses, respectively. Typically, it is therefore substantially smaller than the mass term $m^2_+$ for the chirons $\beta^{\mu\nu}_+$. On the other hand, the effective mixing between $\beta^+_+$ and $W^+$ is enhanced by a factor $-f_tm_t/(f_bm_b)$ as compared to the $\beta^+_--W^+$-mixing, as shown in the comparison of the two graphs in fig. \ref{light-fig6}. 

\noindent
This enhancement is very substantial and may overcompensate the suppression  $\sim \tilde m^2/(m^2_+-m^2_-)$. In this case the diagram in Fig. \ref{light-fig5} dominates the effective chiral vertex between the $W$-boson and the muon. As a result, the $W$-contribution to $\kappa_\mu$ is multiplied by a factor 
\be\label{z6a}
c_{\beta,W}\approx 1+f_tm_t\tilde m^2/\big[f_bm_b(m^2_+-m^2_-)\big]
\ee
and can become large. Its sign depends on the sign of $\tilde m^2$ and $f_t/f_b$. 

Similar $\beta^0_+-\beta^0_-$-mixing contributions are present for the effective chiral vertices of the photon and the $Z$-boson. Now $\tilde m^2$ is replaced by an off diagonal mass term $\tilde m^2_0$ for the neutral chiral tensors, and correspondingly for $m^2_+$ and $m^2_-$. For the photon one also has an additional factor $(-2)$ reflecting the different electric charges of the top and bottom quarks \cite{CW3}. 

\begin{figure}[h!tb]
\centering
\includegraphics[scale=1.2]{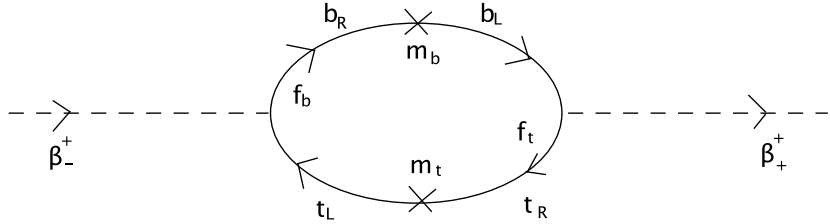}
\caption{Loop contribution to charged chiron mixing}
\label{light-fig7}
\end{figure}

The size of $\tilde m^2$ and $\tilde m^2_0$ could be quite different. This may be seen by observing that we can write down a one loop expression for $\tilde m^2$, shown in fig. \ref{light-fig7}. No such graph exists for the neutral chirons. We recall, however, that the effective chiron propagator is a non-perturbative quantity such that loop-considerations give at best some qualitative insight.

Our discussion shows that a negative value of $\kappa_\mu$ seems well possible. This would imply $\Delta(g-2)>0$. Unfortunately, this investigation has also revealed a substantial uncertainty in the mixing coefficients $c_{\beta,W}$ etc. The corresponding uncertainty in the absolute size of $\kappa_\mu$ directly affects the estimates of $m^2_-(0)$ \eqref{a5b} and $f_\mu f_b$ \eqref{a5c}. For $\kappa_\mu\approx-100$ we would arrive at $m_-{(0)}\approx 4TeV,f_\mu f_b\approx0.2$. Clearly, further quantitative precision needs a better understanding of the chiron propagator. 

In conclusion, we have established the mechanism how the masses of the light generations of quarks and charged leptons are generated within the scenario of chiral freedom. With minor modifications our calculations can be used to compute $m_s/m_b$ or $m_c/m_t$ in terms of $f_s$ and $f_c$. Due to the linearity in the chiral couplings one finds for the first generation the relations
\be\label{z9a}
\frac{m_e}{m_\mu}=\frac{f_e}{f_\mu}~,~\frac{m_u}{m_c}=\frac{f_u}{f_c}~,~\frac{m_d}{m_s}=\frac{f_d}{f_s}.
\ee
The mechanism for the mass generation of the $b$-quark (and probably also the tau-lepton) is different. Even in presence of a nonvanishing top quark mass it requires the spontaneous breaking of the discrete symmetry.

Despite substantial non-perturbative uncertainties concerning the chiron propagator or, correspondingly, the effective chiral vertex of the gauge bosons, we have gained first estimates of the size of the chiral couplings, in particular $f_\mu f_b$. This helps to relate observable quantities, like the anomalous magnetic moment of the muon, to the overall size of the chiron propagator or the chiron mass $m_-$. Knowledge of the latter is crucial for the estimate of possible new LHC-physics arising from the chirons. At very low momenta we have found a rather large value of $m_-\gtrsim 10 TeV$. If the chiron mass term is only moderately momentum dependent the direct chiron effects for the LHC would be rather small - the chiron mass would remain above the momentum range reached by the LHC. In this case the dominant signature of chiral freedom for the LHC would be the non-perturbative effective Higgs-sector, with typical masses for the composite scalar fields in the $500$ GeV range.


\begin{thebibliography}{--}
\bibitem{CW1}C. Wetterich, arXiv: hep-ph/0503164
\bibitem{CW3}C. Wetterich, Phys. Rev. {\bf D74} (2006) 095009, arXiv: hep-ph/0607051
\bibitem{CW2}C. Wetterich, arXiv: hep-th/0509210
\bibitem{CH}M. V. Chizhov, Mod. Phys. Lett. {\bf A8}, 2753 (1993)
\bibitem{CKM}M. Kobayashi, T. Maskawa, Prog. Theor. Phys. {\bf 49}, 652 (1973);\\
N. Cabibbo, Phys. Rev. Lett. {\bf 10}, 531 (1963)
\bibitem{MEX}G. W. Bennett et al., Phys. Rev. Lett. {\bf 92}, 161802 (2004)
\bibitem{MTH}J. P. Miller, E. de Rafael, B. L. Roberts, Rep. Prog. Phys. {\bf 70}, 795 (2007)
\end{thebibliography}
\end{document}